\documentclass[preprint2,longabstract]{aastex}

\usepackage{natbib}

\newcommand{\teff}{$T_{\rm eff}$}
\newcommand{\logg}{log\,$g$}
\newcommand{\he}{HE~1327$-$2326}
\newcommand{\hi}{H\,{\sc i}}
\newcommand{\cai}{Ca\,{\sc i}}
\newcommand{\caii}{Ca\,{\sc ii}}
\newcommand{\caiii}{Ca\,{\sc i/ii}}
\newcommand{\lii}{Li\,{\sc i}}
\newcommand{\eu}[5]{\mbox{$#1\,^#2{\rm #3}^{#4}_{\rm #5}$}}
\newcommand{\eps}[1]{\log \varepsilon({\rm #1})}

\shorttitle{The Subgiant HE 1327$-$2326 and Atomic Diffusion}
\shortauthors{Korn et al.}

\begin{document}


\title{HE~1327-2326, an unevolved star with \boldmath$\mbox{[Fe/H]}<-5.0$\\
    III. Does Its Atmosphere Reflect Its Natal Composition?}

\author{A.J. Korn}
\affil{Department of Physics and Astronomy, Division of Astronomy and Space Physics, Uppsala University, Box 515, 75120 Uppsala, Sweden}
\email{andreas.korn@fysast.uu.se}

\author{O. Richard}
\affil{GRAAL, Universit\'{e} Montpellier II, CNRS, Place E. Bataillon, 34095 Montpellier, France}

\author{L. Mashonkina}
\affil{Institute of Astronomy, Russian Academy of Science, Pyatnitskaya 48, 119017 Moscow, Russia\\
Institut f\"ur Astronomie und Astrophysik der Universit\"at M\"unchen, Scheinerstr. 1, 81679 M\"unchen, Germany}

\author{M.S. Bessell}
\affil{RSAA, The Australian National University, Mt Stromlo, Cotter Rd, Weston, ACT 2611, Australia}

\author{A. Frebel}
\affil{McDonald Observatory, University of Texas, 1 University
Station, C1400, Austin, TX 78712-0259}

\and

\author{W. Aoki}
\affil{National Astronomical Observatory of Japan, Mitaka, Tokyo 181-8588, Japan}



\begin{abstract}
Based on spectroscopic constraints derived from NLTE line formation, we explore the likely range of stellar parameters (\teff\ and \logg) for the hyper-metal-poor (HMP) star \he. Combining the constraints from Balmer line profiles and the \caiii\ ionization equilibrium, a subgiant stage of evolution is indicated. This result is further supported by spectrophotometric observations of the Balmer jump. If a higher \teff\ value was used (as favoured by some photometric calibrations), the spectroscopic analysis would indicate a turnoff-point stage of evolution.

Using a stellar-structure code that treats the effects of atomic diffusion throughout the star in detail, we evolve a low-mass model star to reach the HR-diagram position of \he\ after roughly 13\,Gyr. While the surface abundances are modified significantly (by more than 1\,dex for the case of uninhibited diffusion), such corrections can not resolve the discrepancy between the abundance inferred from the non-detection of the \lii\ resonance line at 6707\,\AA\ and the WMAP-based primordial lithium abundance. As there are numerous processes that can destroy lithium, any cosmological interpretation of a lower-than-expected lithium abundance at the lowest metallicities will have to await sample sizes of unevolved hyper-metal-poor stars that are one order of magnitude larger. The situation remains equally inconclusive concerning atomic-diffusion corrections. Here, attempts have to be made to better constrain internal mixing processes, both observationally and by means of sophisticated modelling. With constraints on additional mixing processes taken from a recent globular-cluster study, the likeliest scenario is that \he's surface abundances have undergone mild depletion (of order 0.2\,dex).
\end{abstract}



\section{Introduction}
The most metal-poor stars provide us with chemical information about the early Universe and the onset of the cosmic chemical evolution. This is based on the assumption that the surface abundances of these old stars indeed reflect the composition of their birth cloud, and have not changed during the long lifetimes. In principle, there are several internal and external processes which may have an impact on the surface composition of old stars, such as internal pollution of the outer layers by nuclear burning products from the stellar interior (in the case the star is sufficiently evolved) or accretion of interstellar gas during $\sim 13$\,Gyr of orbiting though the Galactic potential (the latter process was recently shown to be rather inefficient, see \cite{2008MNRAS.tmpL.137F}).

\he\ is a low-mass star with extreme properties: its excessively low metallicity in terms of iron ([Fe/H]\,$\approx\,-5.6$, \citet{Frebel_etal_2005}) is accompanied by large (3\,--\,4\,dex) over-abundances of the elements C, N and O relative to iron, a property shared by the other two stars known in this metallicity regime (\citet{2002Natur.419..904C}, \cite{2007ApJ...670..774N}). What these stars can tell us about the properties of the first generation of stars (metal-free Population III) is still debated. The abundance pattern has been interpreted to reflect the nucleosynthetic yields of a single Pop\,III supernova (\cite{2005Sci...309..451I,2006A&A...447..623M}), but see \cite{2008ApJ...677..572V} for an alternative scenario potentially explaining the abundance peculiarities of hyper-metal-poor (HMP) stars.

Apart from these interpretative uncertainties, our ability to read the chemical fingerprints encoded in the spectra of these stars is limited by the technical challenge to properly model the line-formation process taking place in their atmospheres. For the most part, the chemical abundances have been inferred using one-dimensional (1D) hydrostatic model atmospheres in local thermodynamic equilibrium (LTE) with rough (estimated) corrections for departures from LTE (so-called NLTE effects). In the case of calcium, the only element observed in two ionization stages, 1D LTE leads to a clear mismatch for the abundances derived from \cai\ and \caii: the abundance from the \cai\ resonance line at 4226\,\AA\ gives an abundance 0.57\,dex below the abundance derived from the \caii\ K line at 3934\,\AA\ \citep{2006ApJ...639..897A}. This may reflect the shortcomings of classical 1D LTE analyses when applied to HMP stars. Explorations using three-dimensional (3D) hydrodynamic model atmospheres have modified the abundance ratios (bringing down the large over-abundances by typically 0.7\,dex for the CNO elements), but have been unable to resolve the above-mentioned discrepancy (reported by \cite{2008ApJ...684..588F} to be 0.46\,dex in 1D-LTE and 0.47\,dex in 3D-LTE).

The LTE assumption is particularly questionable when analysing (hyper-)metal-poor stars, as their atmospheres are UV-transparent leading to strong photo-ionization. Furthermore, electronic collisions are less efficient due to the reduced number of electron donors like Fe and Mg. Both these effects will drive the statistical equilibrium away from LTE. It is thus worthwhile to check whether NLTE can solve the obvious mismatch for the case of calcium.

\he\ is either an evolved main-sequence (MS) or early subgiant (SG) star at \teff\,$\approx$\,6200\,K, i.e., \he\ is a star on the Spite plateau of lithium \citep{Spite_Spite_1982} formed by stars with $T_{\rm eff} \geq 5800$\,K and [Fe/H]\,$\leq$\,$-1.3$ \citep{Charbonnel_Primas_2005}. Despite the uncertainty in the evolutionary status of \he, the relatively unevolved nature of this old star makes it an ideal target for a Li measurement. Other unevolved metal-poor stars have been used to infer the primordial Li abundance. However, the Li doublet at 6707\,\AA\ could not be detected in \he. The upper limit\footnote{The customary logarithmic 12-scale is used here: $\log\,\varepsilon({\rm X}) = \log\,n_{\rm X}/n_{\rm H} + 12$} is $\log\,\varepsilon ({\rm Li})\,\leq\,0.7$, in contrast to the Spite-plateau value of $\log \varepsilon\,({\rm Li})\,\approx\,2.3$ found in other stars. This finding currently lacks a satisfying explanation.

In search for a solution, one can question the assumption that the surface abundances of old stars are constant over time. Even for single MS or SG stars, processes may be at work which systematically alter the chemical surface composition. For solar-type stars, stellar-structure models allowing the atoms to move freely inside the star under the prevailing forces predict sizeable effects in particular for old metal-poor stars. Such processes (gravitational settling and radiative levitation, to name two prominent ones) are collectively referred to as atomic diffusion and require billions of years to alter the abundances of solar-type stars. Convection can suppress their efficiency, but atomic diffusion seems capable of depleting the thin low-mass convective envelopes of metal-poor stars appreciably (see e.g. \cite{Richard_etal_2002}), unless extra mixing processes stir up the boundary layer between the convective and the radiative zone.

Observational evidence in favour of atomic diffusion in the presence of some extra mixing was recently presented \citep{Korn_etal_2006,2007ApJ...671..402K,2008A&A...490..777L} for stars in NGC 6397 ([Fe/H]\,$\approx\,-2$). Based on observed abundance trends between stars from the turn-off point (TOP) to the red-giant branch (RGB), the mixing efficiency needed in the stellar-structure models was empirically determined. This mixing falls in the range expected from observational constraints set by the Spite plateau \citep{Richard_etal_2005}; surface-abundance alterations due to atomic diffusion are then limited to $\pm\,0.3$\,dex and help to harmonize the inferred lithium abundances with the predictions from WMAP-based Big-Bang nucleosynthesis calculations \citep{Korn_etal_2006}.

In this article, we thus explore to which extent the surface abundances of \he\ may be affected by atomic diffusion. After describing the observational material (Sect.\ 2) we derive stellar parameters using spectrophotometry and NLTE diagnostics (Sect.\ 3). Based on these, we present upper limits on the abundance variations induced by atomic diffusion and discuss their consequences (Sect.\ 4). In Sect.\ 5, we summarize our conclusions.

\section{Observations}
We combine data obtained in 2004 and 2005 with Subaru/HDS and VLT/UVES, respectively. The UVES spectra ($R\,=\,40\,000$, \cite{2006ApJ...638L..17F}) are used for the Balmer lines (H$\alpha$ and H$\beta$) and the Ca IR triplet lines. The HDS spectra ($R\,=\,60\,000$, \cite{2006ApJ...639..897A}) cover Ca K 3933\,\AA\ and the \cai\ resonance line 4226\,\AA. In all cases, the signal-to-noise ratio exceeds 200:1. The limiting factor for an abundance analysis of \he\ is thus not to be sought in the quality of the input data.

Spectrophotometric observations of \he\ were obtained with the Double Beam Spectrograph (DBS) on the 2.3m telescope at Siding Springs Observatory, Australia. In June 2006, a ´600B' spectrum from 3200\,\AA\ to 5200\,\AA\ was obtained. In January 2009, a ´300B' spectrum was taken covering the wavelength range from 3200\,\AA\ to 6200\,\AA.
The spectra were taken with the atmospheric dispersion along the slit to ensure that the relative absolute fluxes were accurately measured. The spectra were corrected for continuous atmospheric extinction and wavelength calibrated
using Ne-Ar arc exposures. They were also divided by a continuous white dwarf spectrum to remove the spectrograph and dichroic response as described by \cite{1999PASP..111.1426B,2007PASP..119..605B}.
The spectra were then scrunched and flux corrected using more than five spectrophotometric standards selected each night from the Next Generation Stellar Library STIS observations of Michael Gregg
\footnote{http://www.stsci.edu/hst/HST\textunderscore overview/documents/\linebreak
calworkshop/workshop2005/papers/gregg.pdf}
as re-reduced by Heap and Lindler
\footnote{http://archive.stsci.edu/prepds/stisngsl/index.html}
. The relative absolute fluxes of the NGSL stars are precise to within two percent. The 300B and 600B spectra show excellent agreement.

\section{Stellar parameters}
A crucial ingredient in this investigation is the knowledge of \he's evolutionary status. So far, it has not been possible to distinguish whether \he\ is an MS or an SG star. Luckily, this has no significant impact on the abundances derived from minority species (like \cai\ and Fe\,{\sc i}) and only a small one for majority ones (like \caii, $\approx 0.2$\,dex). For C, N and O, however, there are differences of up to $\approx\,0.5$\,dex between the two cases, as these abundances are exclusively derived from molecular features. Since the CNO abundances with respect to iron are extremely high in this star, these differences have not greatly hampered the interpretation of the overall abundance pattern. But with increasing sophistication of the modelling it is crucial to determine the evolutionary status of this star. This is also the case when comparing to predictions from stellar-structure models.

\subsection{Effective temperature}

From four broad-band indices calibrated on the infrared-flux method (IRFM) of \cite{Alonso_etal_1996}, a mean effective temperature of \teff\,=\,6180\,K was derived \citep{Frebel_etal_2005,2006ApJ...639..897A}. Balmer lines in LTE seemed to favour a somewhat lower effective temperature, \teff\,$\approx$\,6000\,K, irrespective of the broadening theory applied (see Table 7 of \cite{2006ApJ...639..897A}).

We have performed NLTE calculations for \hi\ using the method described in \cite{2008A&A...478..529M}. In brief, the model atom includes levels with principal quantum
numbers up to $n$\,=\,19. Collisional rates include inelastic collisions with electrons and hydrogen atoms. For the latter, we use the formula of \cite{Steenbock_Holweger_1984}. Since it provides only an order of magnitude estimate, we apply it with a scaling factor S$_{\rm H}$ = 2 as favoured by the consistency of the effective temperatures derived from two Balmer lines, H$\alpha$ and H$\beta$, in four metal-poor ([Fe/H] $< -2$) stars \citep{2008A&A...478..529M}.

\begin{figure*}
\includegraphics[scale=1.]{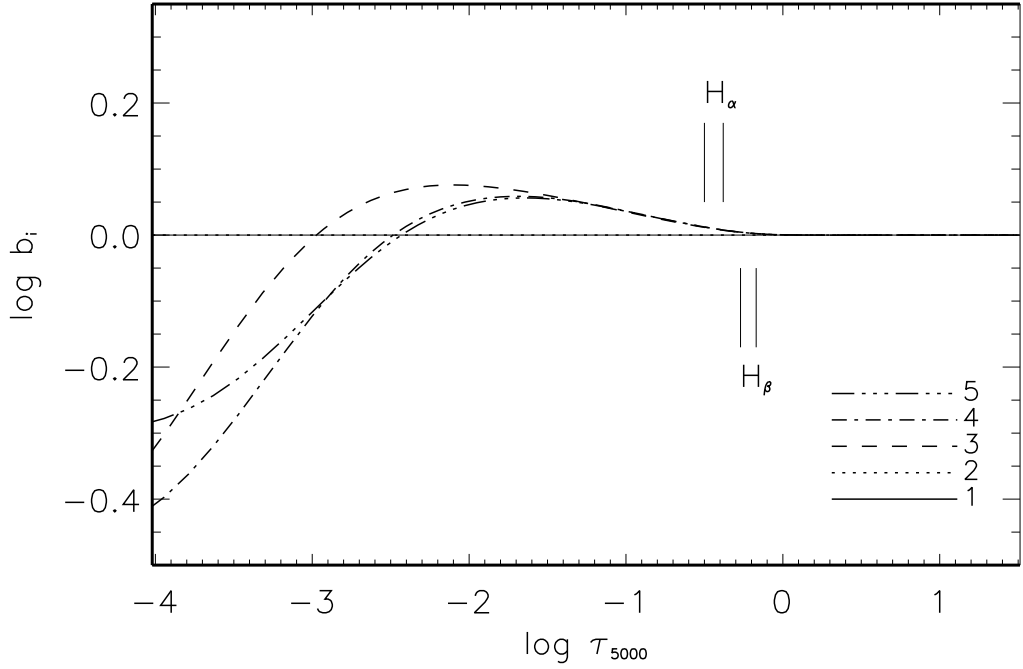}
\caption[]{Departure coefficients $\log b_i$ for the five lowest
levels of \hi\ in the model atmosphere 6180/3.7/$-$4.34. Tick marks indicate the locations of core-to-wing transition formation depths for H$_\alpha$ and H$_\beta$ (0.85 to 0.95 in normalized flux).} \label{h_bf}
\end{figure*}

For the model atmospheres investigated in this study, the ground state and the $n = 2$ level keep their thermodynamic-equilibrium level populations throughout the atmosphere. Fig. \ref{h_bf} shows the behaviour of the departure coefficients $b_i = n_i^{\rm NLTE}/n_i^{\rm LTE}$ of the relevant levels of \hi\ as a function of continuum optical depth $\tau_{5000}$ referring to $\lambda = 5000$\AA\ in the model with \teff/\logg/[Fe/H] of 6180/3.7/$-$4.34. Here, $n_i^{\rm NLTE}$ and $n_i^{\rm LTE}$ are the statistical-equilibrium and thermal (Saha-Boltzmann) number densities, respectively.
Departures from LTE for the $n = 3$ level are controlled by H$\alpha$. In the layers where the continuum optical depth drops below unity H$\alpha$ serves as the pumping transition resulting in an over-population of the upper level. For instance, $b(n = 3) \approx 1.08$ around $\log \tau_{5000} = -0.5$ where the core-to-wing transition is formed. Starting from $\log \tau_{5000} \approx -3$ and further out, photon losses from the wings of H$_\alpha$ cause an under-population of the $n = 3$ level.
We obtain weak NLTE effects for the H$\beta$ profile beyond the core (which is not considered in the analysis), independent of the applied scaling factor for hydrogen collisions.  For H$\alpha$, NLTE leads to a weakening of the core-to-wing transition compared to the LTE case. Exactly this part of profile is sensitive to \teff\ variations.

Table 1 summarizes our \teff\ determination. In NLTE, both H$\alpha$ and H$\beta$ indicate an effective temperature close to \teff\,=\,6120\,K (see Figs.~2 and 3) putting the new spectroscopic effective temperatures in much better agreement with the photometric estimates. For reasons of consistency with previous publications in this series, we adopt \teff\,=\,6180\,K as our current best estimate (a significantly higher effective temperature seems possible using different calibrations of the IRFM \citep{Ramirez_Melendez_2005} and we briefly discuss the impact of this scenario below).

\subsection{Surface gravity}

\begin{figure*}
\includegraphics[angle=90,scale=.50]{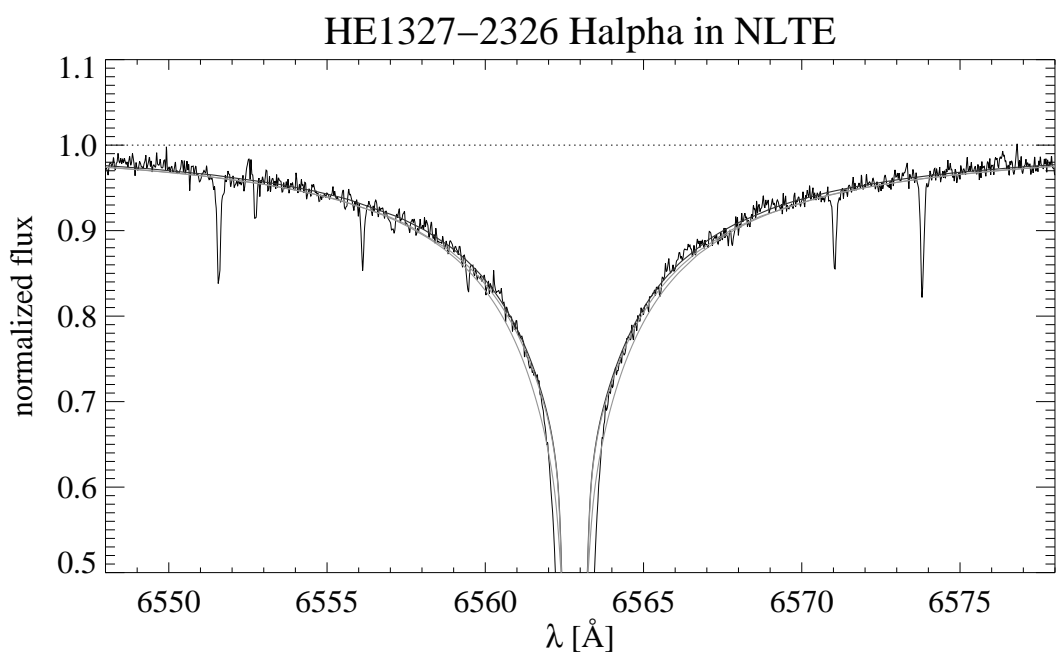}
\caption{Fit to the observed H$\alpha$ profile. An effective temperature of \teff\,=\,6120\,K is indicated in NLTE. The predicted NLTE profile for \teff\,=\,6180\,K is indicated by the upper grey line, the LTE one by the lower one.}
\end{figure*}

\begin{figure*}
\includegraphics[angle=90,scale=.50]{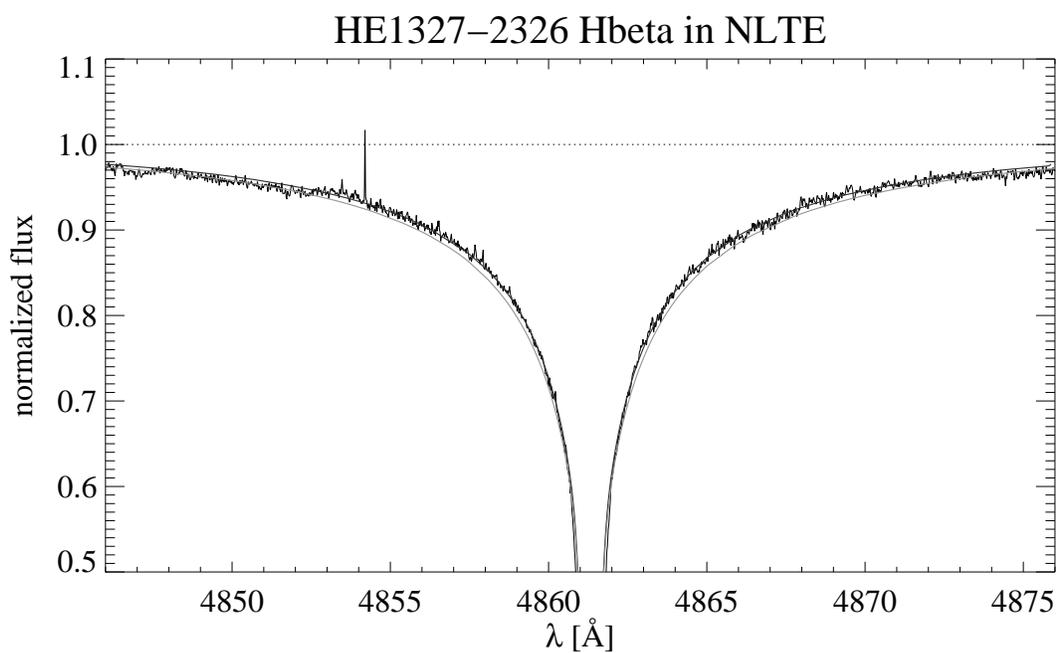}
\caption{Fit to the observed H$\beta$ profile (black line). An effective temperature of \teff\,=\,6120\,K is indicated in NLTE. The predicted NLTE profile for \teff\,=\,6180\,K is indicated by the grey line.}
\end{figure*}

\begin{table*}
\begin{center}
\caption{Spectroscopic effective temperatures as derived from Balmer lines in NLTE.}
\begin{tabular}{lccc}
\tableline\tableline
star & H$\alpha_{\rm NLTE}$ & H$\beta_{\rm NLTE}$ & error\tablenotemark{a}\\
\tableline
\he & 6120\,K & 6120\,K & $\pm$150\,K\\
Sun & 5780\,K & 5780\,K & $+20$\,K, $-60$\,K\\
\tableline
\end{tabular}
\tablenotetext{a}{estimated total error budget including observational uncertainties}
\end{center}
\end{table*}

\begin{figure*}
\includegraphics[angle=0,scale=1.]{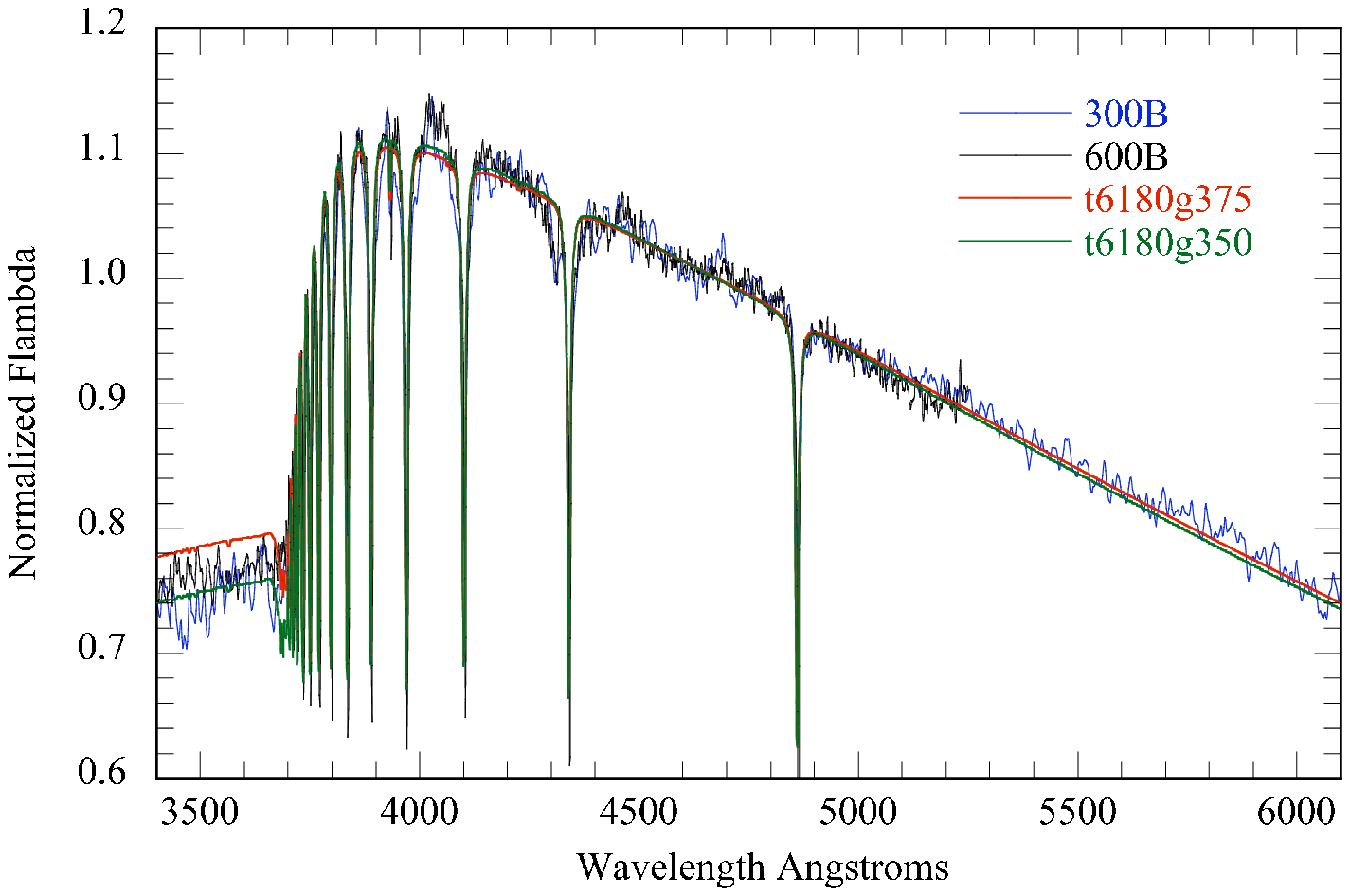}
\caption{Comparison between spectrophotometric observations of \he\ and model-atmosphere predictions (\teff\,=\,6180\,K, [Fe/H]\,=\,$-5$, $E(B-V)\,=\,0.075$). The strength of the Balmer jump indicates a surface gravity of \logg\,=\,3.65.}
\end{figure*}

\begin{figure*}
\includegraphics[scale=1.]{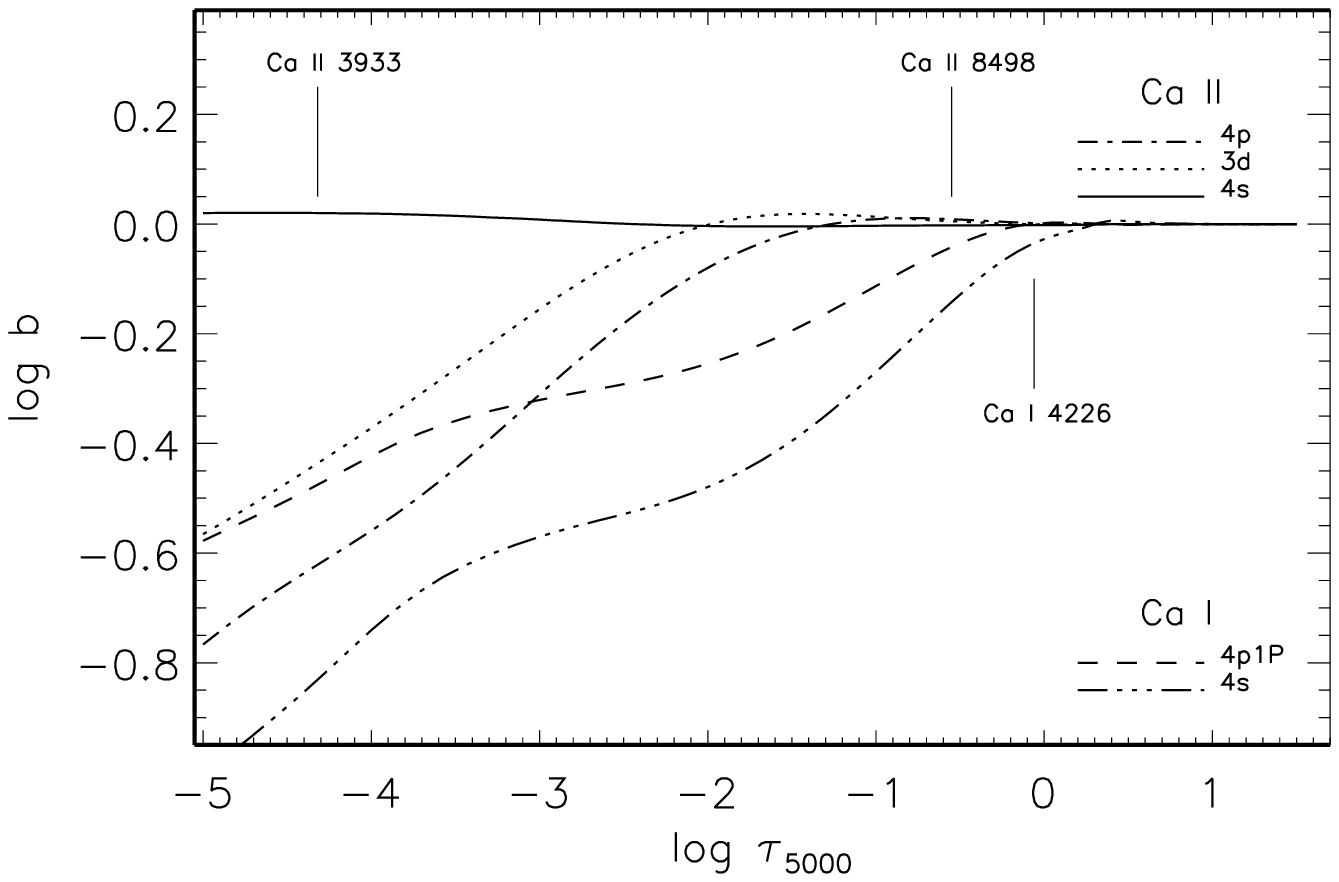}
\caption{Departure coefficients $\log b_i$ for selected
levels of \cai\ and \caii\ in the model atmosphere 6180/3.7/$-$4.34. Tick marks indicate the locations of line-centre optical depth unity for \cai\ 4226\,\AA,  \caii\ 3933\,\AA\ and \caii\ 8498\,\AA.}
\label{ca_bf}
\end{figure*}

The surface gravity of hyper-metal-poor stars is, generally speaking, difficult to constrain observationally. The Balmer jump can serve as a photometric surface-gravity indicator (see, e.g., \cite{2008arXiv0812.0388O}). Figure 4 shows comparisons between the two DBS observations and model fluxes of two MARCS models \citep{2008A&A...486..951G} with \teff\,=\,6180\,K, [Fe/H]\,=\,$-5$ and \logg\ between 3.5 and 3.75 ($E(B-V)\,=\,0.075$ was assumed). As the recent `300B' spectrum was obtained in bright time, higher weight is given to the `600B' observations. A spectrophotometric surface gravity of close to \logg\,=\,3.65 is derived.

Ionization equilibria are a standard spectroscopic method of constraining \logg. The typical attainable precision is 0.3\,dex when line-to-line scatter is of order 0.1\,dex. The accuracy can be increased by considering NLTE effects. The present investigation of \caiii\ in NLTE is based on our earlier work \citep{Mashonkina_etal_2007} where atomic data and the problems of calcium line formation were considered in detail. The model atom used is up-to-date and includes effective collision strengths for \caii\ transitions from recent $R-$matrix calculations of \cite{2007A&A...469.1203M}. Inelastic collisions with neutral H particles are accounted for using the formula of \cite{Steenbock_Holweger_1984} with a scaling factor $S_{\rm H}$\ =\,0.1 as empirically determined from their different influence on \cai\ and \caii\ lines in Solar and stellar spectra. As in the case of H, such a scaling is warranted, as the \cite{Steenbock_Holweger_1984} recipe yields collisional rates with large systematic uncertainties.

Non-LTE calculations are performed for two MAFAGS \citep{1997A&A...323..909F} model atmospheres with effective temperature \teff\,=\,6180\,K and iron abundance [Fe/H]\,=\,$-$4.34, but with different surface gravities,
\logg\,=\,3.7 and 4.5 (the input calcium abundance is iteratively adjusted to match the derived value). The behaviour of the departure coefficients and the line source functions is similar for both models. In the first place, we are
interested in the behaviour of the lower and upper levels from which the investigated lines, \cai\ 4226\,\AA, \caii\ 3933\,\AA\ and \caii\ 8498\,\AA, arise. Figure 5 shows the departure coefficients of the selected levels
 as a function of continuum optical depth $\tau_{5000}$ in the model 6180/3.7/$-$4.34.

For the minority species \cai, the main NLTE effect is over-ionization. It is caused by superthermal radiation of non-local origin below the thresholds of the \cai\ ground state \eu{4s^2}{1}{S}{}{} ($\lambda_{\rm thr}$\,=\,2028\,\AA) and the low-excitation levels \eu{4p}{3}{P}{\circ}{} ($\lambda_{\rm thr}$\,=\,2930\,\AA) and \eu{3d}{3}{D}{}{} ($\lambda_{\rm thr}$\,=\,3450\,\AA). The \cai\ resonance line is weaker relative to its LTE strength not only due to the general over-ionization but also due to the larger departure coefficient of the upper level compared to that for the lower one resulting in a line source function $S_{lu} \simeq b($\eu{4p}{1}{P}{\circ}{}$)/b($\eu{4s^2}{1}{S}{}{}$)\,B_\nu > B_\nu$ and diminished line absorption. The NLTE abundance correction $\Delta_{\rm NLTE} = \eps{NLTE}-\eps{LTE}$ = +0.31\,dex and +0.25\,dex for the models 6180/3.7/$-$4.34 and 6180/4.5/$-$4.34, respectively.

In the line formation layers, \caii\ dominates the element number density. No process seems to affect the \caii\ ground state population and $4s$ keeps its thermodynamic-equilibrium value. \caii\ resonance-line pumping produces a slightly enhanced excitation of the level $4p$ in the atmospheric layers between $\log\tau_{5000}$\,=\,$-$0.3 and $-$1.3. In the very metal-poor atmosphere of \he, the metastable level $3d$ is only weakly coupled to the ground state and follows $4p$ until photon losses in the \caii\  8498 line start to become important. Departures from LTE for \caii\ 8498 are weak resulting in $\Delta_{\rm NLTE}$\,=\,$-$0.02\,dex and $-$0.01\,dex for the models 6180/3.7/$-$4.34 and 6180/4.5/$-$4.34, respectively. Because of the larger line strength (and thus shallower formation depth) of \caii\ 3933, this line suffers from somewhat larger NLTE effects: $-$0.15\,dex and $-$0.06\,dex for the subgiant and dwarf model, respectively.

\begin{table*}
{\scriptsize
\begin{center}
\caption{Calcium abundances as derived from individual lines. Like in \cite{2006ApJ...639..897A}, a microturbulence value of 1.7\,km/s and 1.5\,km/s was assumed in the line formation calculations for the subgiant and dwarf, respectively.\label{tbl-2}}
\begin{tabular}{lrcccc}
\tableline\tableline
line & $W_\lambda$\tablenotemark{a} & $\eps{Ca}$ in NLTE & $\eps{Ca}$ in LTE & $\eps{Ca}$ in NLTE & $\eps{Ca}$ in LTE \\
 & & @ \logg\,=\,3.7 &  @ \logg\,=\,3.7 &  @ \logg\,=\,4.5 & @ \logg\,=\,4.5\\
\tableline
3933\,{\AA} & 128.9\,m{\AA} & 1.17 & 1.32 & 1.29 & 1.35\tablenotemark{b}\\
4226\,{\AA} & 2.7\,m{\AA} & 1.22 & 0.91 & 1.15 & 0.90\\
8498\,{\AA} & 9.6\,m{\AA} & 1.27 & 1.29 & 1.56 & 1.57\\
\tableline
all lines $\pm$ 1$\sigma$ & & 1.17\tablenotemark{c} $\pm$ 0.05 & 1.17 $\pm$ 0.23 & 1.33\tablenotemark{d} $\pm$ 0.21 & 1.27 $\pm$ 0.34 \\
\tableline
\end{tabular}
\tablenotetext{a}{as determined from the the theoretical spectra after profile fitting}
\tablenotetext{b}{In this analysis, the fit is based on the wings, as the wings and core cannot be fitted simultaneously.}
\tablenotetext{c}{corresponds to [Ca/Fe]\,=\,0.47 on the scale of \cite{2006ApJ...639..897A}}
\tablenotetext{d}{corresponds to [Ca/Fe]\,=\,0.62 on the scale of \cite{2006ApJ...639..897A}}
\end{center}
}
\end{table*}

We note that NLTE effects for \caiii\ in 3D line formation calculations are currently unexplored. For the minority species Li\,{\sc i} in metal-poor stellar atmospheres, \cite{Barklem_etal_2003} concluded that 3D\,+\,NLTE is not greatly different from 1D\,+\,NLTE with 1D\,+\,NLTE results overestimating the Li abundance by up to 0.1\,dex. Similar effects can be expected for \cai. For \caii, the departures from LTE are likely to be small, similar to the 1D case.

Table 2 summarizes the results concerning calcium. Good agreement (within 0.05\,dex) is obtained for the three different lines when NLTE and a surface gravity of \logg\,=\,3.7 are used (numerically, a slightly lower \logg\ value is indicated which we disregard for reasons explained in Sect.~\ref{AD}). At the dwarf-like surface gravity, the \logg-sensitive line \caii\ 8498 is more than 0.4\,dex discrepant from \cai\ 4226. From this and the line-to-line scatter we derive an uncertainty for \logg\ of 0.2\,dex.

A rigourous NLTE modelling of \caiii\ in 1D is thus capable of fully accounting for the \caiii\ abundance discrepancy seen in previous analyses. \he\ is therefore a subgiant, a conclusion in accord with indications from Balmer lines (see \cite{2006ApJ...639..897A}). The [Ca/Fe] ratio is +0.47\,dex which is not an unusual value for a halo star. {\he\ is a benchmark case for future NLTE studies for \caiii\ in 3D model atmospheres.}

We estimate the effects of crucial atomic data on the accuracy of the NLTE Ca abundances derived from \cai\ 4226 and \caii\ 8498 (the two most critical lines for our purpose) using the model 6180/3.7/$-$4.34. The largest effects are found upon disregarding hydrogenic collisions altogether. But even in that case, none of the line abundances changes by more than 0.1\,dex and our conclusion about \he\ being a subgiant holds.

\section{Stellar-structure model predictions}\label{AD}

\subsection{Model description}

The stellar evolutionary models used here are calculated with the physics of particle transport that is known from first principles. In radiative zones this implies taking into account atomic diffusion including gravitational settling, thermal diffusion and radiative accelerations, in addition to the purely diffusive term. The diffusion velocity of each species is computed with the equations as developed in Burgers (1969) which take into account the interactions between all diffusing species.

The detailed treatment of atomic diffusion is described in \cite{TuRiMiIgRo98}. In these models, the Rosseland opacity and radiative accelerations are computed in each layer for the exact local chemical composition using OPAL monochromatic opacities and for each time step during the evolution. The radiative accelerations are from \cite{RiMiRoetal98} with corrections for redistribution from \cite{GoLeArMi95} and \cite{LeMiRi2000}.
Convection and semi-convection are modelled as diffusion processes as described in \cite{RiMiTu2000} and
\cite{RiMiRi2001}.

In one set of models, additional mixing is included in the diffusion equation by adding an effective diffusion coefficient as in \cite{RicherMiTu2000} and \cite{RiMiRi2001}. The parameters specifying this effective diffusion coefficients are indicated in the name assigned to the model, which are the same as those used in the \cite{2007ApJ...671..402K} study on NGC 6397. This additional mixing is introduced in an ad-hoc manner: here no attempt is made to connect it to physical processes like rotation. As shown by \cite{RicherMiTu2000}, the most important factor is the radial extent of the mixed layers. All models considered here were assumed to be chemically homogeneous on the pre-main sequence with Solar relative mass fractions as given in Table~\ref{tab:Xinit}. To mimic the abundance peculiarities of \he, the relative concentrations of the CNO elements were increased by +4\,dex compared with the scaled Solar mixture. As the most abundant metals in \he, they have structural effects on the models.

\subsection{Best-fitting models}

\begin{figure*}
\includegraphics[scale=.70]{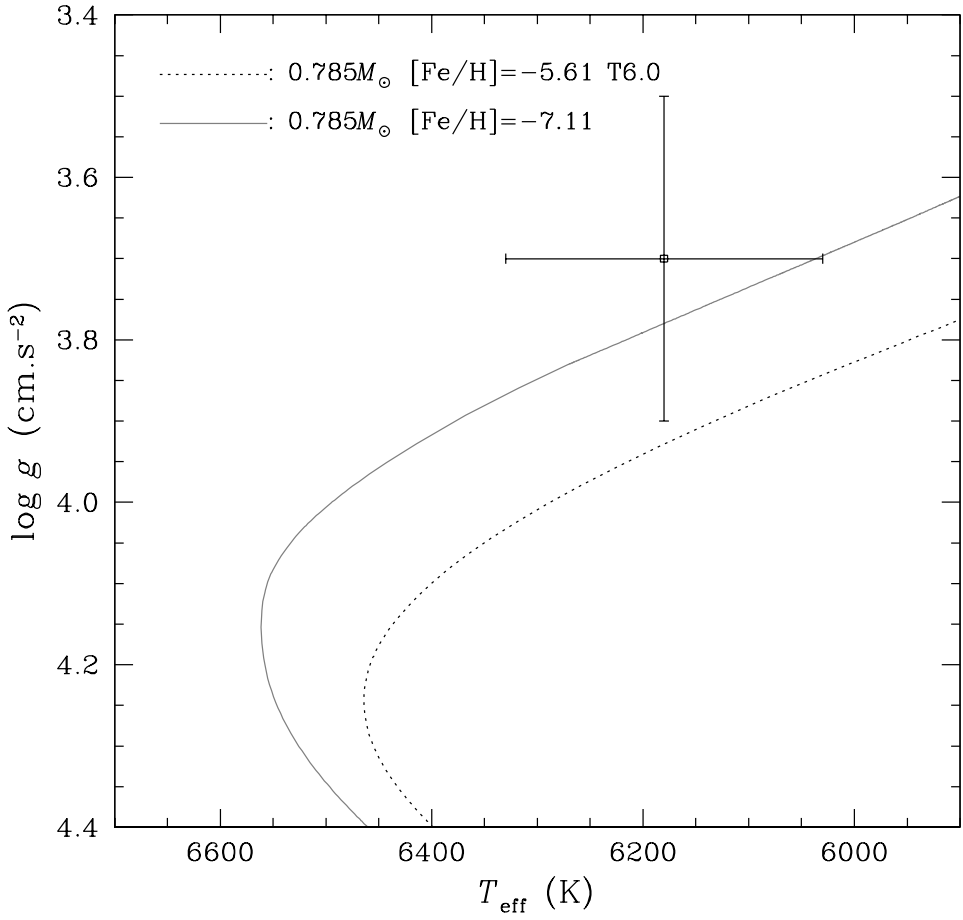}
\caption{Position of \he\ relative to evolutionary models constructed from the two diffusion models.}
\end{figure*}

\begin{deluxetable}{lcccc}
\tabletypesize{\scriptsize}
\tablewidth{.75\textwidth}
\tablecaption{Initial chemical composition of the two different diffusion models. The abundance variations after $\approx$ 13.3\,Gyr can be estimated from Figure \ref{abundancevariations}}
\tablecolumns{7}
\tablehead{\colhead{Element} & {\tiny Note} & \multicolumn{2}{c}{Mass fraction} \\
\cline{3-4} & & [Fe/H]=$-$5.61, T6.0 & [Fe/H]=$-$7.11, diffusive }
\startdata
H  \dotfill & & 7.642$\times 10^{-1}$ & 7.645$\times 10^{-1}$ \\
$^4$He \dotfill & & 2.354$\times 10^{-1}$ & 2.354$\times 10^{-1}$ \\
$^7$Li \dotfill & & 2.0$\times 10^{-9}$ & 2.0$\times 10^{-9}$ \\
$^{12}$C  \dotfill & \tablenotemark{a} & 8.633$\times 10^{-5}$ & 2.763$\times 10^{-6}$ \\
N  \dotfill & \tablenotemark{a} & 2.647$\times 10^{-5}$ & 8.471$\times 10^{-7}$ \\
O  \dotfill & \tablenotemark{a} & 2.409$\times 10^{-4}$ & 7.708$\times 10^{-6}$ \\
Ne \dotfill & & 4.927$\times 10^{-9}$ & 1.577$\times 10^{-10}$ \\
Na \dotfill & & 9.989$\times 10^{-11}$ & 3.197$\times 10^{-12}$ \\
Mg \dotfill & & 1.875$\times 10^{-9}$ & 6.002$\times 10^{-11}$ \\
Al \dotfill & & 1.623$\times 10^{-10}$ &  5.194$\times 10^{-12}$ \\
Si \dotfill & & 2.023$\times 10^{-9}$ & 6.473$\times 10^{-11}$ \\
P  \dotfill & & 1.748$\times 10^{-11}$ & 5.594$\times 10^{-13}$ \\
S  \dotfill & & 1.056$\times 10^{-9}$ & 3.380$\times 10^{-11}$ \\
Cl \dotfill & & 2.248$\times 10^{-11}$ & 7.192$\times 10^{-13}$ \\
Ar \dotfill & & 2.697$\times 10^{-10}$ & 8.631$\times 10^{-12}$ \\
K  \dotfill & & 9.989$\times 10^{-12}$ & 3.197$\times 10^{-13}$ \\
Ca \dotfill & & 1.873$\times 10^{-10}$ &  5.994$\times 10^{-12}$ \\
Ti \dotfill & & 9.989$\times 10^{-12}$ & 3.197$\times 10^{-13}$ \\
Cr \dotfill & & 4.994$\times 10^{-11}$ & 1.598$\times 10^{-12}$ \\
Mn \dotfill & & 2.747$\times 10^{-11}$ & 8.791$\times 10^{-13}$ \\
Fe \dotfill & & 3.586$\times 10^{-9}$ & 1.147$\times 10^{-10}$ \\
Ni \dotfill & & 2.223$\times 10^{-10}$ & 7.112$\times 10^{-12}$ \\
\cline{1-5}
Z  \dotfill & & 3.545$\times 10^{-4}$ & 1.135$\times 10^{-5}$ \\
\enddata
\tablenotetext{a}{increased by +4\,dex relative to the assumed solar mixture}
\label{tab:Xinit}
\end{deluxetable}

\begin{figure*}
\vspace*{-4cm}
\hspace*{-3cm}
\includegraphics[scale=.90,clip=]{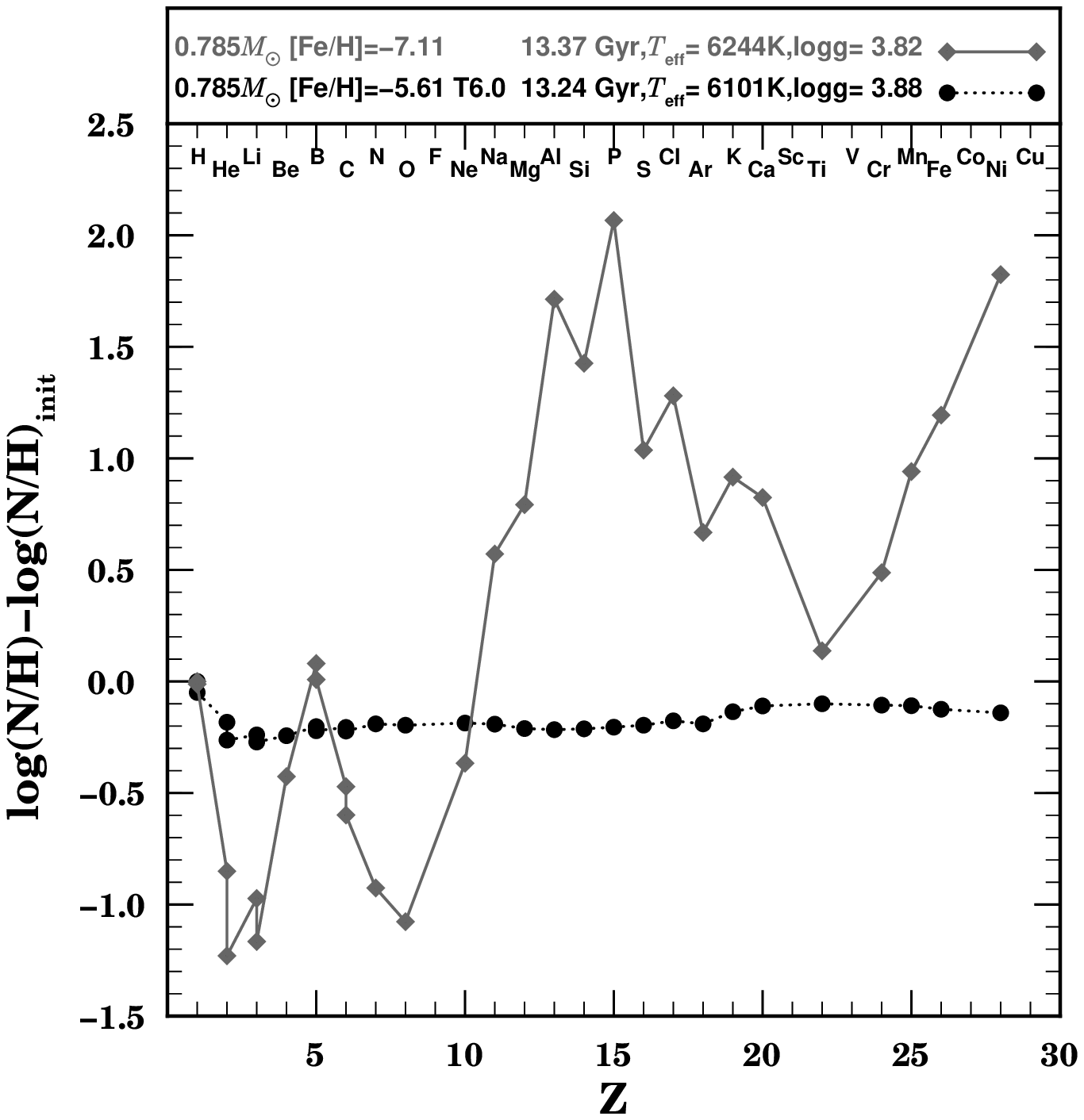}
\vspace*{-2cm}
\caption{Surface-abundance variations as predicted by the two diffusion models, relative to the initial abundances given in Table 3.}
\label{abundancevariations}
\end{figure*}

We computed a variety of models with masses around $0.78\,$M$_{\odot}$ for this study, both purely diffusive ones and ones including additional mixing. The mass and initial composition of each model was varied to roughly meet the observational constraints of \he: the surface abundance of iron should be within a factor of two of [Fe/H]\,=\,$-5.8$ (the average of the 1D and 3D LTE abundances derived by \cite{2006ApJ...639..897A} and \cite{2008ApJ...684..588F}) when the model star reaches the error ellipse of \he\ on the subgiant branch after roughly 13\,Gyr (see Figure 6).

Two models of 0.785\,M$_{\odot}$ fulfilled these a-posteriori requirements. The first one has an initial [Fe/H] of [Fe/H]$_i$\,=\,$-7.11$ and was computed with atomic diffusion and radiative accelerations, but without additional transport processes in the radiative zone. As shown by \citet{RiMiRi2002III} for higher metallicities (between $-2.31$ and $-4.31$), such a star is expected to display strong surface abundance variations close to and past the TOP. The second one ([Fe/H]\,=\,$-5.61$) additionally includes additional transport as needed to explain abundance variations of NGC~6397 \citep{Korn_etal_2006,2007ApJ...671..402K}.

\subsection{Birth-cloud vs.~present-day abundances}

As can be seen from Figure 7, abundance corrections are large for the model with uninhibited diffusion: after $\approx$ 13.4\,Gyr the $M\,=\,0.785\,$M$_\odot$ [Fe/H]$_i$\,=\,$-7.11$ model has reached the subgiant branch and its surface lithium is depleted by almost 1.2\,dex. Simultaneously, radiative levitation has increased its convection-zone abundances of iron by 1.2\,dex. Other elements, e.g. aluminium and nickel, are affected by an even larger amount. We compared the diffusion-corrected abundance pattern of \he\ to that of HE 0107$-$5240 and find that the abundance patterns of the two stars do not become more similar. When additional mixing is introduced into the model, the surface-abundance changes are largely moderated: helium and lithium only settle by $\approx$\,0.3\,dex (incidentally, the same amount as predicted for NGC 6397, \cite{2007ApJ...671..402K}), iron by only $\approx$\,0.1\,dex.

If \he\ turned out to be significantly hotter (\teff\,=\,6450\,K), then a surface gravity of \logg\,$\approx$\,4.0 would be derived in the framework of this analysis. From the stellar-structure point of view, this is a meaningful combination of stellar parameters (see Figure 6), albeit not supported by our understanding of the formation of Balmer lines. The same diffusion models could be used to predict corrections to the surface abundances and these would turn out to be quite similar within a given set of model assumptions. This is owed to the fact that the change in stellar parameters effectively moves the star along its fast early post-TOP evolutionary track.

Given our present ignorance of the extent of the additional mixing (and how it scales with, e.g., metallicity), it is difficult to draw firm conclusions about which of the two diffusion scenarios is more realistic. It is conceivable that the mixing efficiency declines towards lower metallicities, as the convective envelope decreases in extent and mass. Nonetheless, even in the absence of mixing the correction is not large enough to reconcile the upper limit on the lithium abundance ($\eps{Li}\,\leq\,0.7$) with the WMAP-based primordial lithium-abundance prediction of Big-Bang nucleosynthesis ($\eps{Li}\,=\,2.63$, \cite{Spergel_etal_2007}). If the low lithium content of \he\ turned out to be the rule among unevolved HMP star, then this may tell us something specific about the environment these stars were able to form in. More work is needed to identify similar stars in the outer halo.

Until better constraints can be set on the mixing efficiency, we recommend to use the abundance corrections based on the model with extra mixing. These corrections are moderate and generally negative. The birth cloud from which \he\ formed was somewhat more metal-rich than the surface abundances of \he\ show at the present time.

As discussed in \cite{2007ApJ...671..402K}), helium diffusion has a structural effect on the atmosphere of a star and  can be mapped as a shift in the surface-gravity scale. The gravitational settling of helium entails a shift of the surface gravity of $\approx\,+0.05$\,dex. This correction was considered when deriving our best current estimate of \logg\,=\,3.7\,$\pm$\,0.2.

\section{Conclusions}
We have derived spectroscopic stellar parameters for \he\ using 1D NLTE techniques: Balmer lines profiles in NLTE yield \teff\,=\,6120 $\pm$ 150\,K confirming the photometrically determined effective temperature of \teff\,=\,6180\,K. The calcium ionization equilibrium in NLTE points towards a subgiant surface gravity of \logg\,=\,3.7 $\pm$ 0.2, as does spectrophotometry of the Balmer jump. Using an effective temperature of 6450\,K (as favoured by a revision of the IRFM, \cite{Ramirez_Melendez_2005}), results in a surface gravity typical for turn-off stars. While this is also a sensible stellar-parameter combination from the point of view of stellar evolution and the \caiii\ ionization equilibrium, it finds no support from Balmer-line analyses.

The exploratory investigations into how much \he's surface abundances may be affected by atomic diffusion show that large (more than 1\,dex) corrections are in principle possible. The material from which \he\ formed would then have been more metal-poor ([Fe/H]$_i$\,$\approx\,-7$) and significantly more lithium-rich ($\eps{Li}_i\,\leq\,1.9$).  However, there is no compelling evidence in favour of uninhibited diffusion in stars that have convective envelopes. If additional mixing below the convective envelope is postulated (as needed to explain abundance trends in NGC 6397, \cite{2007ApJ...671..402K}), then the abundance corrections are limited to a factor of two. Assuming the additional-mixing efficiency to be independent of metallicity, the corrections are thus moderate given the uncertainties associated with present-day 1D-(N)LTE or 3D-LTE abundances of \he\ and other HMP stars.

It is too early to draw firm conclusions about our ability to reliably read the chemical fingerprints in the atmospheres of stars like \he\ and their nucleosynthetic significance as such. More such stars need to be identified in deep wide-angle surveys to leave the realm of low-number statistics. Efforts in this direction are under way. Systematic comparisons between chemical abundances in HMP dwarfs, subgiants and giants may then reveal the efficiency of mixing moderating atomic diffusion. Simultaneously, the modelling has to be improved. This will mean replacing the hydrostatic approximation with full radiation hydrodynamics, both in stellar-structure and stellar-atmosphere modelling. A similar transition is required for photospheric line-formation calculations in terms of statistical equilibria rather than LTE, as documented by the successes of this NLTE study.

\acknowledgements
A. J. K. acknowledges a research fellowship by the Swedish Research Council (Vetenskapr{\aa}det). O.~R. thanks the Centre Informatique National de l'Enseignement Sup\'{e}rieur (CINES) and the R\'{e}seau Qu\'{e}b\'{e}cois de Calcul de Haute Performance (RQCHP) for providing the computational resources required for this work. M.~L. is supported by Presidium RAS Programme ``Origin and evolution of stars and the Galaxy'' and the Deutsche Forschungsgemeinschaft with grant GE 490/34-1. A.~F. is supported through the W.~J.~McDonald Fellowship of the McDonald Observatory. W.~A. is supported by a Grant-in-Aid for Science Research from JSPS (grant 18104003).


\end{document}